\documentclass{egpubl}
\usepackage{eurovis2025}

\EuroVisShort  % for EuroVis additional material
\usepackage[T1]{fontenc}
\usepackage{dfadobe}

\usepackage{cite}  % comment out for biblatex with backend=biber
\BibtexOrBiblatex
\electronicVersion
\PrintedOrElectronic
\ifpdf \usepackage[pdftex]{graphicx} \pdfcompresslevel=9
\else \usepackage[dvips]{graphicx} \fi

\usepackage{egweblnk}
\usepackage{xcolor}
\usepackage{amsmath}
\title[Grid Labeling: Crowdsourcing Task-Specific Importance from Visualizations]%
      {Grid Labeling: Crowdsourcing Task-Specific Importance from Visualizations}

\author[M. Chang]
{\parbox{\textwidth}{
    \centering 
    Minsuk Chang$^{1}$\orcid{0009-0007-5088-8991},
    Yao Wang$^{2}$\orcid{0000-0002-3633-8623},
    Huichen Will Wang$^{3}$\orcid{0009-0007-5941-4047}, 
    Andreas Bulling$^{2}$\orcid{0000-0001-6317-7303}, and
    Cindy Xiong Bearfield$^{1}$\orcid{0000-0002-1451-4083}
    }
\\
{\parbox{\textwidth}{\centering 
    $^1$Georgia Institute of Technology, USA\\
    $^2$University of Stuttgart, Germany\\
    $^3$University of Washington, USA
}}}
\begin{document}

% uncomment for using teaser
\teaser{
 \includegraphics[width=\linewidth]{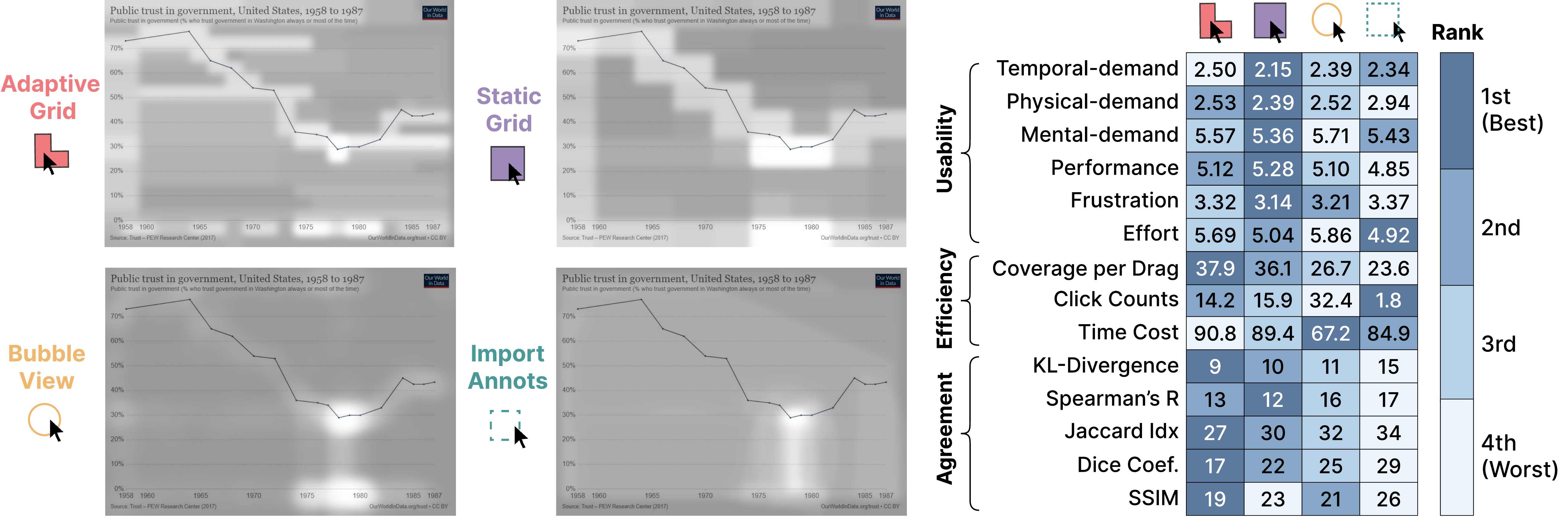}
 \centering
  \caption{(left) Example importance annotations for the task “Between which two years was public trust in government the lowest?” obtained using three tools: Adaptive/Static Grid (Ours), BubbleView\cite{bubbleView}, and ImportAnnots\cite{importAnnot}. (right) Average quantity and ranking for each annotation method for all metrics used in our experiment. A darker color represents a higher ranking in each row.}
\label{fig:teaser}
}

\maketitle
%-------------------------------------------------------------------------
\begin{abstract}
Knowing where people look in visualizations is key to effective design. Yet, existing research primarily focuses on free-viewing-based saliency models--- although visual attention is inherently task-dependent. 
Collecting task-relevant importance data remains a resource-intensive challenge.
To address this, we introduce \textit{Grid Labeling} -- a novel annotation method for collecting task-specific importance data to enhance saliency prediction models. 
Grid Labeling dynamically segments visualizations into Adaptive Grids, enabling efficient, low-effort annotation while adapting to visualization structure.
We conducted a human-subject study comparing Grid Labeling with existing annotation methods, ImportAnnots, and BubbleView across multiple metrics. 
Results show that Grid Labeling produces the least noisy data and the highest inter-participant agreement with fewer participants while requiring less physical (e.g., clicks/mouse movements) and cognitive effort. An interactive demo and the accompanying dataset are available at \href{https://github.com/jangsus1/Grid-Labeling}{\texttt{https://github.com/jangsus1/Grid-Labeling}}.


\end{abstract}  
%-------------------------------------------------------------------------

\section{Introduction}

Where do people look in visualizations under tasks?
Understanding salient parts of visualizations is crucial for designing compelling visualizations that optimally support analytic tasks~\cite{graphicDesignImportance}. 
However, modeling saliency is challenging as where people visually focus is inherently task-dependent. 
For example, during free viewing, participants may primarily engage in bottom-up processes driven by visually salient elements, such as a bright red patch~\cite{kinchla1979ordervisualprocesse}.
However, when given an analytic task, participants are more likely to engage in top-down processing, directing their attention toward task-relevant visualization regions. For example, in a ``find the extreme'' task, they may focus on the tip of the highest bar~\cite{gilbert2013topdown}.
In this context, task-dependent saliency is strongly related to importance, which involves actively filtering areas with sufficient information for task-solving.
Existing work has defined the image's ``importance'' as regions where individual annotators believe as important~\cite{importAnnot}. 
We supplement this definition by taking a task-dependent approach to propose a new alternate definition for ``task-specific importance'' as ``\textit{the minimum area in a visualization required for a user to complete a task successfully}.'' 

However, existing mouse-tracking-based saliency collection methods rely on free-viewing~\cite{turkeyes, importAnnot, bubbleView}, as they are designed to encourage users to explore and describe an image. However, because visualizations are often used for analytic tasks~\cite{amar2005low}, we posit that an effective model for predicting where people look should consider task relevance when identifying regions of importance.
To address this limitation, we contribute \textit{Grid Labeling}, a toolkit to enhance existing saliency prediction models with task-specific annotation. 
A key advantage of our tool is that it is more resource-efficient than traditional methods, such as eye-tracking or mouse-tracking~\cite{graphicDesignImportance, importAnnot, bubbleView, turkeyes, salchartQA}.

Grid Labeling segments visualizations into Adaptive Grids that dynamically adjust based on existing graphical elements, making it easily adapted to various visualization sizes and designs. 
This approach enables participants to identify critical areas simply by clicking relevant grids, eliminating the need for cumbersome mouse interactions, such as free-form drawing to annotate regions~\cite{importAnnot} or clicking the same regions multiple times~\cite{bubbleView}.
Moreover, Grid Labeling streamlines data collection, reducing the number of participants required to converge to a stable importance map.
In a human-subject experiment, we demonstrate that, compared to the two popular approaches, ImportAnnots~\cite{importAnnot} and BubbleView~\cite{bubbleView}, Grid Labeling produces less noisy data with higher levels of agreement between participant responses.
Additionally, participants reported lower perceived effort when using our method. We also explore the key distinction between saliency and importance, contributing to differences in annotation duration.

The specific contributions of our work are three-fold:

\begin{itemize}
    \item We introduce the Grid Labeling method for capturing ``task-specific importance'' in visualizations, which enhances task-specific saliency modeling.

    \item Through a human-subject study, we illustrate the importance of considering task-specific ``areas of importance'' in visualizations.

    \item We quantitatively demonstrate that our Grid Labeling method outperforms traditional crowd-sourcing methods for collecting task-specific importance data in visualizations. Participants in our study could identify important areas with less effort and with higher inter-participant agreement.
\end{itemize}

\section{Related Work}
Researchers have been leveraging eye tracking methodologies from human perception research to model how people perceive images~\cite{shanmuga2015eye, conklin2016using}.
These models help assess the appearance and salience of visual representations, enabling eye movement tracking to understand the perceptual and cognitive mechanisms of scene perception~\cite{itti1998model} and object detection~\cite{borji2015salient}.
The existing saliency models perform well in naturalistic scenes; however, there are unique perception rules and cognitive biases in the artificial world of data visualization~\cite{correll2012comparing}, and, thus, these models do not accurately predict where people would look in visualizations. 
Visualization researchers have been building visual saliency models geared to visualizations~\cite{DVSaliencyModel2017Matzen, bylinskii2016should}.
However, these models rely on handcrafted features, making it difficult to generalize to complex visualizations. Additionally, these models cannot incorporate textual information to generate task-specific saliency maps since the prediction is solely based on visual inputs.

With deep learning, gaze data became the ground truth for saliency models~\cite{scannerDeeply, scanpath}, increasing prediction performance and enabling task-specific saliency~\cite{salchartQA}. 
These models require large datasets, but collecting accurate gaze data is expensive and cumbersome. 
To address this, researchers proposed gaze proxies. For example, WebGaze~\cite{webgaze} offers low-cost webcam-based data collection for online studies, though it struggles with quality due to low resolution and poor calibration.
Therefore, mouse cursor-based annotation tools~\cite{jiang2015salicon,bubbleView,importAnnot} were proposed to improve data quality. Among these methods, BubbleView~\cite{bubbleView} was the most widely used tool for capturing visual saliency and importance~\cite{graphicDesignImportance, salchartQA}.
However, BubbleView is primarily designed for exploring images and gathering information, which differs slightly from the goal of capturing perceived importance. As a result, while BubbleView is well-suited for measuring visual saliency, it may not be the best tool for capturing 
task-specific importance~\cite{turkeyes}. Built upon these prior approaches' limitations, our Grid Labeling aims to collect responses that cover all essential areas of the visualization with minimum noise, leading to more efficient data collection.

\section{Grid Labeling}
\begin{figure}[t]
    \centering
    \includegraphics[width=0.95\linewidth]{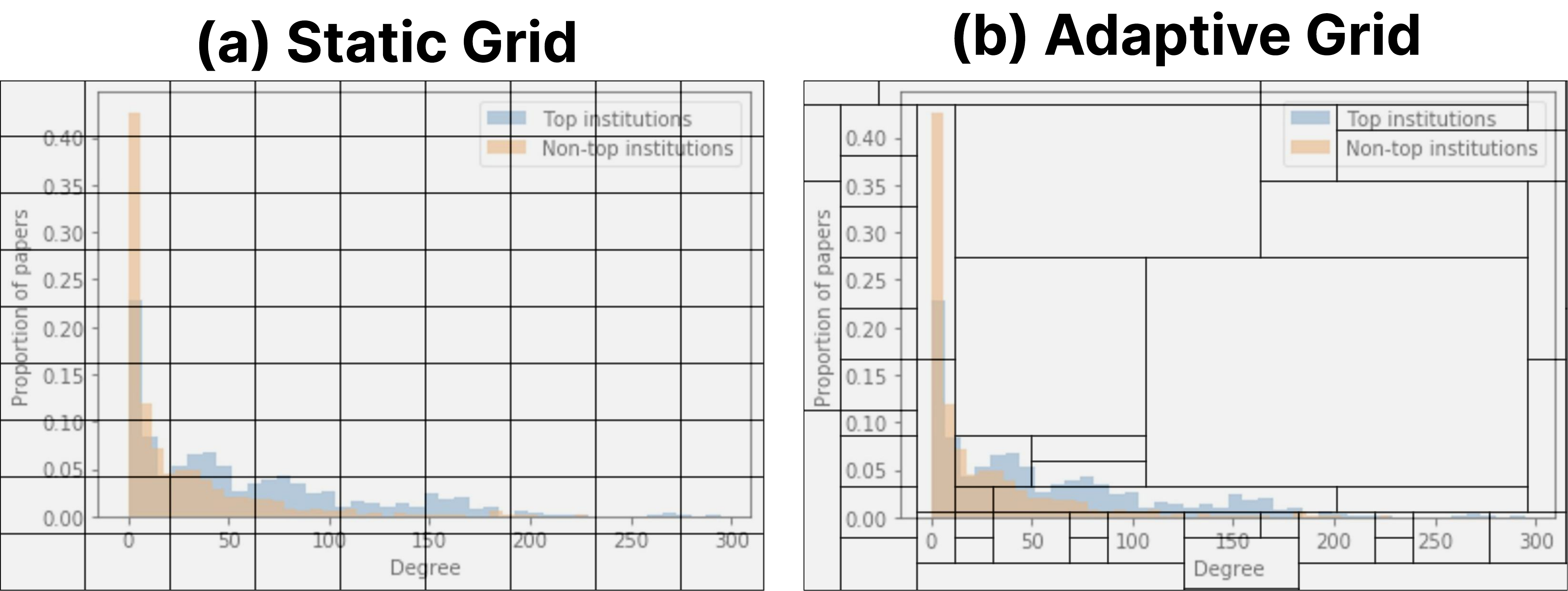}
    \caption{Comparison of Static and Adaptive Grid segmentations applied to a histogram from CharXiv~\cite{charxiv}. Static Grid splits the visualization into equal-sized rectangles, while Adaptive Grid dynamically produces patches that fit the visualization layout.}
    \label{fig:grids}
    \vspace{-8mm}
\end{figure}

Existing saliency and importance annotation methods use circular~\cite{bubbleView} or freeform~\cite{importAnnot} shapes, which do not preserve the structure of visual elements, particularly when creating saliency maps by Gaussian kernels~\cite{scanpath}.
Inspired by Google's reCaptcha~\cite{recaptcha}, we propose \textbf{Grid Labeling}, a patch-based annotation approach that addresses this limitation. With Grid Labeling, users annotate specific areas by clicking on image patches (\autoref{fig:grids}). This binary interaction\,---\,clicking or not clicking\,---\,enforces discrete annotations, facilitating faster response aggregation by promoting higher consensus. 
\subsection{Static Grid} \label{grids:static}
As a baseline, we first designed the Static Grid by dividing the visualization’s height and width into N equal sections, resulting in $N^2$ patches. We leave N as a hyperparameter, which was set to 8 in our experiment. This made the patch size approximately equivalent to the recommended circle size in BubbleView~\cite{bubbleView}. 

\subsection{Adaptive Grid} \label{grids:adaptive}
To further reduce annotation time and effort, we introduce an Adaptive Grid that groups smaller \textit{tiles} into larger \textit{blocks} aligned with the visualization’s layout.

\noindent\textbf{Step 1: Split Regions.}
We divide the visualization into three regions: text, edge, and background. We filter out the text area with PaddlePaddle OCR, followed by the Canny edge detection algorithm to extract graphical elements. The remaining tiles not identified as text or edges are labeled as background.

\noindent\textbf{Step 2: Defining the Tile Space.}
Let's assume we are filling in the visualization with small tiles with the size of $t$ (e.g., 32px), which is the minimum patch size. 
Then the image would be covered with a grid of tiles with dimensions $M \times N$ ($M=\left\lfloor \text{Height} / t \right\rfloor, N=\left\lfloor \text{Width} / t \right\rfloor$), forming $G \in \{0,1\}^{M \times N}$. 
We individually build the binary grid for each region (text, edge, or background), which will be covered with larger blocks in Step 3. 
Each entry $G_{i,j}$ is set to 1 if tile $(i,j)$ belongs to the selected region and 0 otherwise. 

\noindent\textbf{Step 3: Optimizing Block Arrangement.}
We then assign larger rectangular blocks that can minimally cover the entire grid. Define binary decision variables $B_{i,j}^{w,h} \in \{0,1\}$, where $B_{i,j}^{w,h} = 1$ indicates that a rectangular block of size $w \times h$ tiles is placed with its top-left corner at tile $(i,j)$. Our objective is to minimize the total number of blocks:
$\min \sum_{i,j,w,h} B_{i,j}^{w,h}$
, while the entire region must be covered once without overlap between the blocks. This merges coverage and non-overlap requirements into a single constraint:
\[
\sum_{\substack{(i',j',w,h) \\ i' \leq i < i'+h,\; j' \leq j < j'+w}} B_{i',j'}^{w,h} \;=\; G_{i,j}, 
\quad \forall (i,j).
\]
We solve this optimization problem using Integer Linear Programming (ILP) with OR-Tools’ constraint programming solver. By enforcing the exact coverage of each tile, we get patches that cover the visualization while respecting the background contents.
\section{User Study}

We investigate how crowdworkers annotate important areas in visualizations with different annotation methods. We first demonstrate that annotations differ when users are instructed to annotate task-specific vs. task-agnostic areas, motivating the need for more task-specific annotation data to be collected. We then compare four methods participants can use to identify minimal task-relevant areas in a visualization: BubbleView, ImportAnnots, Static Grid, and Adaptive Grid. We evaluate these methods based on four metrics: task completion time, the number of participants required for convergence, annotation effort (e.g., number of clicks), and usability.

\subsection{Participants and Design}

We conducted a power analysis based on pilot results. Considering the smallest effect size across all comparison metrics (Cohen's $f$ = 0.2715 for cognitive load), a target sample of 152 participants would yield 80\% power to detect an overall difference between annotation methodologies at an $\alpha$ level of 0.05. 
We recruited participants from Prolific ($M_{age}$ = 38.5, $\sigma = 11.9$, 52 females) and compensated them \$12 per hour. 
We curated a set of 18 visualizations from ChartQA~\cite{chartQA} and CharXiv~\cite{charxiv}, covering a diverse range of chart types (e.g., bar, stacked bar, pie, line, Choropleth map, heatmap, histogram, scatterplot, and contour plots). Each participant was then shown these selected charts in a random order. 
In a between-subject set-up, participants were randomly assigned to annotate them via one of four tools: BubbleView, ImportAnnots, and Static/Adaptive Grid.

\subsection{Procedure}
The study was conducted as a between-subject experiment. After consenting to the experiment, participants were instructed on how to use the assigned annotation tool (BubbleView, ImportAnnots, Static Grid, or Adaptive Grid).
They first solved an example task with a simple bar chart with one of the following prompts: \textit{annotate the important area and describe key points}, \textit{annotate the area minimally required for you to identify the highest value in the chart} (results see Section~\ref{results:taskbasedagnostic}).
Then, they labeled 18 visualizations using the same tool. 
For ImportAnnots, participants were instructed as \textit{annotate important areas related to answering the question}. 
For BubbleView, they were just asked to answer the question following the prior work's design~\cite{salchartQA}. 
For Grid Labeling (Adaptive/Static), they were instructed to annotate ``Task-Specific Importance.'' 
In the end, they reported the tool's usability using NASA-TLX~\cite{hart1988development}, completed an assessment of their visualization literacy~\cite{mini-vlat}, and provided demographic information.

\subsection{Results: Importance vs. Free-Viewing} \label{results:taskbasedagnostic}

\begin{figure}[ht]
    \centering
    \includegraphics[width=0.9\linewidth]{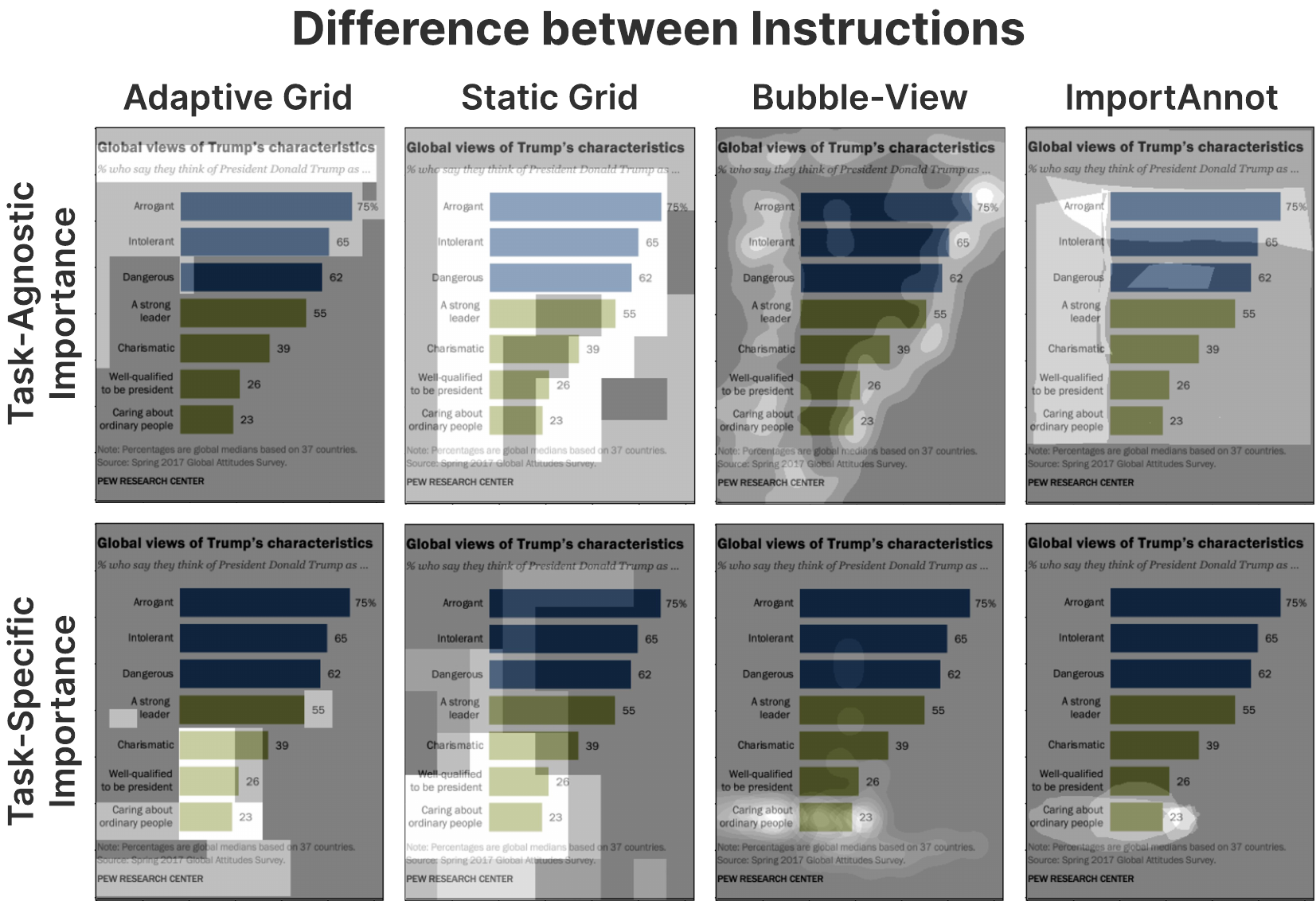}
    \caption{Averaged participants' annotations for four tools when applying different instructions. The task-specific instruction was ``What is the important area regarding this question: What is the minimum value of the bar?''.}
    \label{fig:instruction}
    \vspace{-3mm}
\end{figure}

We demonstrate the participants' annotation behavior appeared significantly different when they were instructed to annotate components of the visualization they found important during free-viewing, compared to when they were instructed to annotate the importance area in response to a specific task, as shown in \autoref{fig:instruction}.
During task-agnostic annotation, the importance area is more evenly distributed across the visualization with a slight emphasis on the top of the visualization and the title text (aligned with existing work such as patterns identified by ~\cite{graphicDesignImportance}).
In contrast, the annotations cluster around the area with the smallest bar at the bottom of the visualization in response to the find minimum task.
This further validates our case that existing models trained on free-viewing annotation and eye-tracking data might not be the most predictive for visualization saliency, considering salient regions can vary with user intent and tasks.

\subsection{Results: Methods Comparison}

\label{results:comparison}
\begin{figure}[t]
    \centering
    \includegraphics[width=0.95\linewidth]{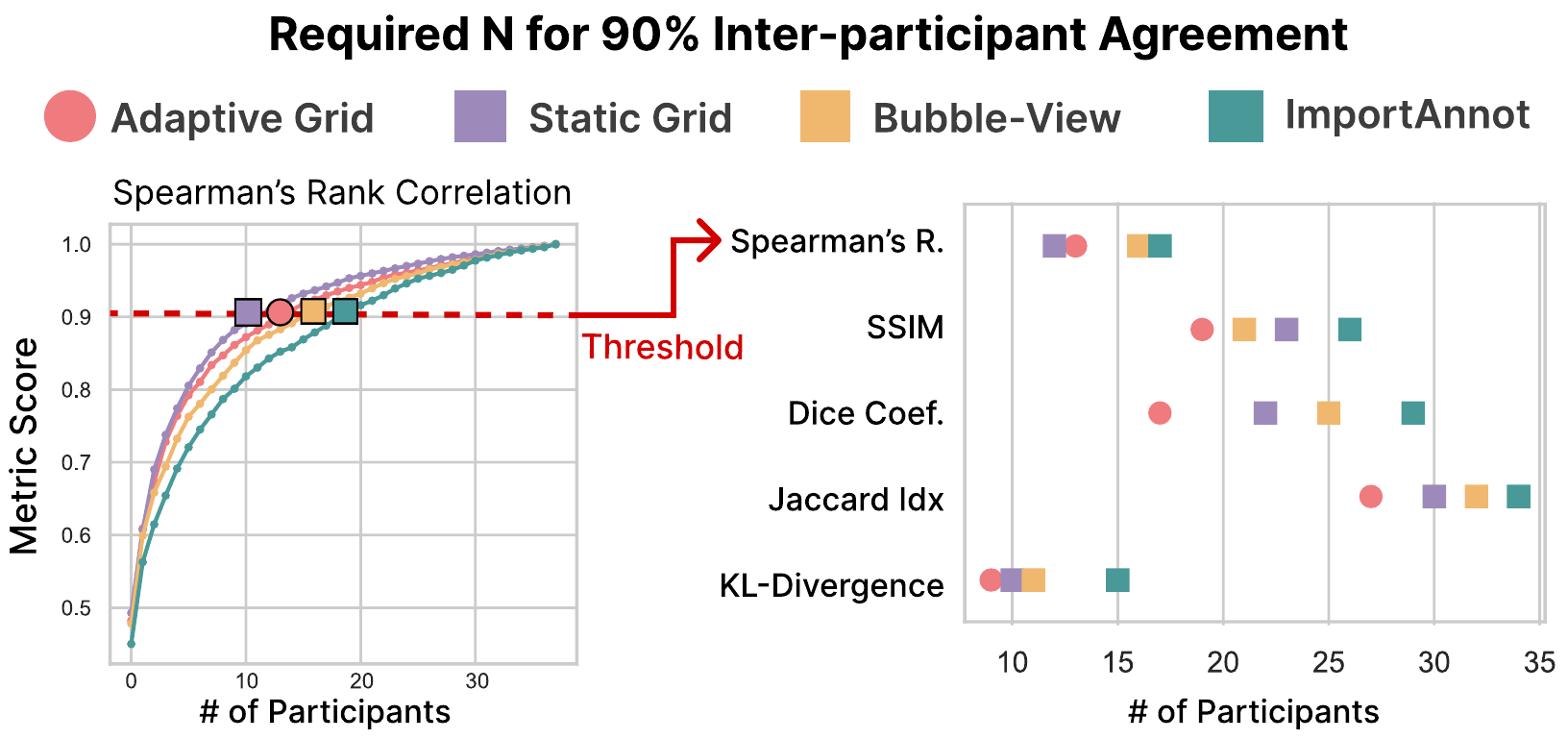}
    \caption{Required number of participants to reach 90\% similarity compared to the aggregated importance. Five different similarity metrics were used to measure the convergence.} 
    \label{fig:convergence}
    \vspace{-5mm}
\end{figure}

In the user study, we compared four annotation methods using five metrics: usability, click counts, annotated area per mouse travel distance, annotation speed, and inter-participant agreement. The group means per metric are summarized in ~\autoref{fig:teaser}.

\noindent\textbf{Usability: Participant Task-Load Analysis}
To measure the usability of the annotation methods, we used the NASA-TLX scale, which has six categories (mental demand, physical demand, temporal demand, effort, performance, and frustration).
The MANOVA suggested that the annotation methods statistically differ (Wilks’ $\Lambda$ = 0.8626, F(18, 7362.88) = 21.94, $p < 0.001$), while independent ANOVAs showed differences except for frustration. 
A post hoc Tukey test followed the five remaining categories.
ImportAnnots showed higher physical demand than others ($p\leq0.001$). Static Grid had lower temporal demand than Adaptive Grid and BubbleView ($p\leq0.023$) while also having higher performance than BubbleView and ImportAnnots ($p\leq0.028$). For effort, ImportAnnots and Static Grid were better than other tools ($p<0.001$).

\noindent\textbf{Interaction Efficiency: Click Count Comparison.}
The ANOVA (F(32,608) = 248.995) and post hoc Tukey test suggested that the significant effect is driven by the difference in click count between all pair-wise comparisons among methods except adaptive and static grid ($p = 0.42$). Participants labeled important areas with adaptive and static grids using fewer clicks than BubbleView. The grid-based methods required more clicks than ImportAnnots, but that is caused by the stroke tool, which allowed participants to annotate a large area while dragging the mouse with a single click. 

\noindent\textbf{Coverage Efficiency: Annotated Area per Mouse Travel Distance.}
We measured the coverage efficiency by dividing the annotated area size by the mouse travel distance during annotation. With ANOVA (F=30.33) and post hoc Tukey ($p<0.001$), we observed a significant difference between Adaptive Grid and Static Grid, having higher coverage efficiency than BubbleView and ImportAnnots.  

\noindent\textbf{Time Efficiency: Annotation Speed Across Methods.}
ANOVA (F=11.79) and the following Tukey test showed BubbleView required less time per annotation compared to other tools ($p<0.001$), demonstrating its efficiency in annotation speed. However, we argue that the underlying reason stems from the difference between saliency and importance, where capturing importance may naturally involve more intention during the annotation process~\cite{turkeyes}.

\noindent\textbf{Convergence Speed: Agreement Across Participants.}
We examined the number of participants needed to achieve 90\% similarity with the aggregated mask using five similarity metrics: Spearman’s Rank Correlation~\cite{spearman}, Structural Similarity Index~\cite{ssim}, Dice Coefficient~\cite{dice}, Jaccard Index~\cite{jaccard}, and Kullback-Leibler (KL) Divergence~\cite{kldivergence}, often used to measure difference between continuous 2d maps. We measured the convergence of 10 different randomized orders of responses for more generalized results with smoother graphs. As shown in \autoref{fig:convergence}, the Adaptive Grid and Static Grid generally converge faster than the other tools across most metrics, while the Adaptive Grid was the best performing in all metrics except Spearman's R.

\section{Discussion \& Future Work}
We contribute Grid Labeling, an annotation method for efficiently crowdsourcing task-dependent important areas of visualizations.
Grid Labeling outperformed other approaches across all metrics, as shown in~\autoref{fig:teaser}.
While ImportAnnots~\cite{importAnnot} had the lowest click count and BubbleView~\cite{bubbleView} required the least time, Adaptive Grid achieved the highest inter-participant agreement with the fewest participants across multiple metrics (e.g., SSIM, Dice, Jaccard, KL). 
Meanwhile, Static Grid demonstrated higher usability, as indicated by NASA-TLX~\cite{hart1988development}.
These results highlight the potential of Grid Labeling in training task-specific saliency models, minimizing text overemphasis, and enhancing predictive accuracy.
Considering the trade-offs, we recommend using an Adaptive Grid for maximizing convergence and a Static Grid to improve usability.

Since the present study did not explore why participants labeled certain grids as important and relied on an inter-participant agreement for quality control, future work could investigate the reasoning behind these selections to provide a more high-level, representational explanation, collecting a large-scale dataset and training a task-specific importance prediction model.
Additionally, future work could refine Adaptive Grid generation using vision-based LLMs to enhance annotation usability by semantically filtering less important visualization grids through an automated pipeline.

\section*{Acknowledgements}
This work was supported by the National Science Foundation under grants IIS-2237585 and IIS-2311575.

% bibtex
\bibliographystyle{eg-alpha-doi} 
\bibliography{references}       

% biblatex with biber
% \printbibliography                

%-------------------------------------------------------------------------
%Color tables are no longer required for purely electronic publications.
\newpage

\end{document}